\documentclass[prd,twocolumn,floatfix,preprintnumbers,showpacs]{revtex4}
\usepackage{graphicx}
\usepackage{dcolumn}
\usepackage{bm}
\usepackage{color}

\def\lsim{\mathrel{\mathop
  {\hbox{\lower0.5ex\hbox{$\sim$}\kern-0.8em\lower-0.7ex\hbox{$<$}}}}}
\def\gsim{\mathrel{\mathop
  {\hbox{\lower0.5ex\hbox{$\sim$}\kern-0.8em\lower-0.7ex\hbox{$>$}}}}}

\begin{document}
\newcommand{\mincir}{\raise
-2.truept\hbox{\rlap{\hbox{$\sim$}}\raise5.truept 
\hbox{$<$}\ }}
\newcommand{\magcir}{\raise
-2.truept\hbox{\rlap{\hbox{$\sim$}}\raise5.truept
\hbox{$>$}\ }}
\newcommand{\minmag}{\raise-2.truept\hbox{\rlap{\hbox{$<$}}\raise
6.truept\hbox
{$>$}\ }}

\newcommand{\1}{$\spadesuit$}
\newcommand{\half}{{1\over2}}
\newcommand{\nad}{n_{\rm ad}}
\newcommand{\niso}{n_{\rm iso}}
\newcommand{\ncor}{n_{\rm cor}}
\newcommand{\fiso}{f_{\rm iso}}
\newcommand{\ii}{\'{\'i}}
\newcommand{\bk}{{\bf k}}
\newcommand{\Ocdm}{\Omega_{\rm cdm}}
\newcommand{\ocdm}{\omega_{\rm cdm}}
\newcommand{\OM}{\Omega_{\rm m}}
\newcommand{\OB}{\Omega_{\rm b}}
\newcommand{\oB}{\omega_{\rm b}}
\newcommand{\OX}{\Omega_{\rm X}}
\newcommand{\cltt}{C_l^{\rm TT}}
\newcommand{\clte}{C_l^{\rm TE}}
\newcommand{\calR}{{\cal R}}
\newcommand{\calS}{{\cal S}}
\newcommand{\Rrad}{{\cal R}_{\rm rad}}
\newcommand{\Srad}{{\cal S}_{\rm rad}}
\newcommand{\calPR}{{\cal P}_{\cal R}}
\newcommand{\calPS}{{\cal P}_{\cal S}}
\newcommand{\etal}{{\it et al.~}}
\newcommand{\lya}{{Lyman-$\alpha$~}}
\input epsf

\preprint{IFT-UAM/CSIC-05-34, LAPTH-1113/05, astro-ph/0509209}
\title{
Squeezing the window on isocurvature modes with the Lyman-$\alpha$ forest}
\author{Mar\'\i a Beltr\'an,$^{1,2}$
Juan Garc{\'\i}a-Bellido,$^1$
Julien Lesgourgues,$^3$
Matteo Viel$^4$} 
\affiliation{
$^1$Departamento de F\'\i sica Te\'orica \ C-XI, Universidad
Aut\'onoma de Madrid, Cantoblanco, 28049 Madrid, Spain\\
$^2$Astronomy Centre, University of Sussex, 
Brighton BN1 9QH, United Kingdom\\
$^3$Laboratoire de Physique Th\'eorique LAPTH, F-74941
Annecy-le-Vieux Cedex, France\\
$^4$Institute of Astronomy, Madingley Road, Cambridge CB3 0HA, United Kingdom
}
\date{September 8, 2005}
\begin{abstract}
Various recent studies proved that cosmological models with a
significant contribution from cold dark matter 
isocurvature perturbations are still
compatible with most recent data on cosmic microwave background 
anisotropies and on the shape
of the galaxy power spectrum, provided that one allows for a very blue
spectrum of primordial entropy fluctuations ($\niso > 2$). However,
such models predict an excess of matter fluctuations on small scales,
typically below $40 \, h^{-1}\mathrm{Mpc}$. We show that the proper
inclusion of high-resolution high signal-to-noise Lyman-$\alpha$
forest data excludes most of these models.
The upper bound on the isocurvature fraction $\alpha=\fiso^2/(1+\fiso^2)$,
defined at the pivot scale $k_0=0.05$~Mpc$^{-1}$, is pushed down to
$\alpha<0.4$, while $\niso=1.9\pm1.0$ (95\% confidence limits).
We also study the bounds on curvaton models characterized by maximal 
correlation between curvature and isocurvature modes, and a unique 
spectral tilt for both. We find that $f_{\rm iso}<0.05$ (95\% c.l.)
in that case. For double inflation models with two
massive inflatons coupled only gravitationally, the
mass ratio should obey $R < 3$ (95\% c.l.).
\end{abstract}

\maketitle
 
\section{Introduction}

With the most recent measurements of the cosmic
microwave background (CMB) anisotropies and large scale structures
(LSS) of the universe as well as various other astronomical
observations, it is now possible to have a clear and consistent
picture of the history and content of the universe since
nucleosynthesis. 
In particular, it is well established that the
cosmological perturbations which gave rise to the CMB anisotropies and
the LSS of the universe were inflationary, with a close to 
scale-invariant Harrison-Zeldovich spectrum. Moreover, the
CMB and LSS data allow to test the paradigm of adiabaticity
of the cosmological perturbations and hence the precise nature of the
mechanism which has generated them.

The simplest realizations of the inflationary paradigm predict an
approximately scale invariant spectrum of adiabatic (AD) and Gaussian
curvature fluctuations, whose amplitude remains constant outside the
horizon, and therefore allows cosmologists to probe directly the
physics of inflation from current CMB and LSS observations.  However,
this is not the only possibility.  Models of inflation with more than
one field generically predict that, together with the adiabatic
component, there should also be entropy, or isocurvature
perturbations~\cite{Linde:1985yf,Polarski:1994rz,Garcia-Bellido:1995qq,Gordon:2000hv,Wands:2002bn,Finelli:2000ya},
associated with fluctuations in number density between different
components of the plasma before photon decoupling, with a possible
statistical correlation between the adiabatic and isocurvature
modes~\cite{Langlois:1999dw}.  Baryon isocurvature (BI) perturbations
and cold dark matter isocurvature (CDI) perturbations were proposed
long ago~\cite{Efstathiou:1986} as an alternative to adiabatic
perturbations. These BI and CDI modes are qualitatively similar, since
they are related by a simple rescaling factor $\Omega_{\rm
B}^2/\Omega_{\rm cdm}^2$, or $\Omega_{\rm B}/\Omega_{\rm cdm}$ for the
cross-correlation: thus, by studying the case of mixed AD + CDI modes,
one implicitly includes the case of AD + BI, for which the allowed
isocurvature fraction is larger roughly by the above factor evaluated
near the maximum likelihood model.  A few years ago, two other modes,
neutrino isocurvature density (NID) and velocity (NIV) perturbations,
have been added to the list~\cite{Bucher:1999re}. Moreover,
isocurvature perturbations have been advocated in order to explain the
high redshift of reionization claimed by the WMAP team
~\cite{zaroubisilk}.

Note, however, that in the case all
fields thermalize at reheating, no isocurvature mode will survive
\cite{Weinberg}. The simplest assumption for generating observable CDI
perturbations is that one of the inflaton fields remains uncoupled
from the rest of the plasma between inflation and its decay into CDM
particles.  Since baryons and neutrinos are usually assumed to be in
thermal equilibrium in the early Universe, it is more difficult to
build realistic models for the generation of BI, NID and NIV modes
than for CDI - but some possibilities still exist, based on non-zero
conserved quantities and chemical
potentials (see e.g.~\cite{Weinberg,LUW,Gordon:2003hw}).

Moreover, it is well known that entropy perturbations seed
curvature perturbations outside the
horizon~\cite{Polarski:1994rz,Garcia-Bellido:1995qq,Gordon:2000hv}, so
that it is possible that a significant component of the observed
adiabatic mode could be maximally correlated with an isocurvature
mode. Such models are generically called {\em curvaton
models}~\cite{Enqvist:2001zp,Lyth:2001nq,Moroi:2001ct,LUW}, 
and are now widely studied as
an alternative to the standard paradigm. Furthermore, isocurvature modes
typically induce non-Gaussian signatures in the spectrum of primordial
perturbations \cite{Bartolo:2004if}.

In the last few years, various models with a correlated mixture of
adiabatic and isocurvature perturbations have been tested by several
authors, with different combinations of data sets and theoretical
priors. A crucial difference between these analyses lies in the
assumptions concerning the scale-dependence of the various modes. Some
groups assumed for simplicity that the adiabatic and isocurvature mode
shared exactly the same scale-dependence
\cite{Bucher:1999re,Trotta:2001yw,Moodley}, but enriched the analysis
by considering the full mixtures of several modes at a time (AD, CDI,
NID, NIV).  Other groups concentrated on the (correlated) mixture of
two modes only (AD+CDI in
\cite{Valiviita:2003ty,Beltran:2004uv,Kurki-Suonio:2004mn}, AD+NID and
AD+NIV in \cite{Beltran:2004uv}), with a different power law for the
three components (adiabatic, isocurvature and cross-correlation), as
expected in the general case. Finally, an intermediate approach
consists in studying the mixture of two modes with a scale-independent
mixing angle, i.e., only two tilts
\cite{Amendola:2001ni,Peiris,Crotty:2003aa,Ferrer:2004nv,Parkinson}.
In addition to these references, some groups studied the case of the
curvaton scenario, which requires some specific analyses
\cite{Gordon:2002gv,Gordon:2003hw,Ferrer:2004nv,Lazarides:2004we}
since it involves a maximal correlation/anti-correlation and a unique
spectral index for the adiabatic and isocurvature modes.  Furthermore,
two groups have quantified the need for isocurvature modes through a
Bayesian Evidence computation on the basis of current CMB and galaxy
power spectrum data, reaching somewhat different conclusions due to
a different choice of priors \cite{Beltran:2005xd,Trotta:2005ar}.

In this work, we are particularly interested in mixed models with
AD+CDI modes and three different tilts, for which it was shown in Refs.
\cite{Beltran:2004uv} and \cite{Kurki-Suonio:2004mn} that a
significant fraction of isocurvature perturbations is still
allowed. This sounds surprising at first sight, since the isocurvature
mode is known for suppressing small-scale CMB anisotropies. This is
true indeed for a scale-invariant spectrum of primordial isocurvature
fluctuations, but not in general: a significant isocurvature
contribution with a very blue tilt ($\niso \sim 3$) can contribute to
CMB anistropies even on small scales, and can be compatible to some
extent with the CMB temperature and temperature-polarization spectra,
in spite of the small shift induced in the scale of the acoustic
peaks. These models predict generically an excess of matter
fluctuations on small scales. Using the shape and amplitude of the
{\it linear} power spectrum derived from galaxy surveys at wavenumbers
$k<0.15 \, h/$Mpc, one can exclude such an excess for wavelengths
$\lambda=2\pi/k$ larger than $40 \, h^{-1}$ Mpc.  The main goal of
this work is to push the constraints down by making use of
Lyman-$\alpha$ forest data, which probe large-scale structure at
redshift $z \sim (2 - 3)$ and on scales $(1-40) \, h^{-1}
\mathrm{Mpc}$, in the mildly non-linear regime. Therefore, in any
comparison between Lyman-$\alpha$ observations and linear theoretical
predictions, it is necessary to take into account the non-linear
evolution with N-body or hydrodynamical simulations.

Usually, these simulations are carried under the assumption of
adiabaticity. However, it is not difficult to generalize them to the
case of mixed adiabatic plus isocurvature models.  During matter
domination, the perturbations seeded by each of the two modes are
indistinguishable: the only difference lies in their scale-dependence,
but not in their nature or time--evolution. So, a given mixed model is
entirely specified by a single matter transfer function, defined for
instance soon after the time of equality. Therefore, the
Lyman-$\alpha$ forest data can be safely applied to non-adiabatic
models provided that one takes into account the fact that the matter
transfer function has more freedom than in the purely adiabatic case.
In the following analysis,
we will carefully take this point into account.

We will use here the linear matter power spectrum inferred from two
large samples of quasar (QSO) absorption spectra \cite{kim04,croft} using
state--of--the--art hydrodynamical simulations \cite{vhs} combined
with cosmic microwave background data from the WMAP
satellite~\cite{WMAP}; as well as from the small-scale temperature
anisotropy probed by VSA~\cite{VSA}, CBI~\cite{CBI} and
ACBAR~\cite{ACBAR}; from the matter power spectrum measured by the
2-degree-Field Galaxy Redshift Survey (2dFGRS)~\cite{2dFGRS} and the
Sloan Digital Sky Survey (SDSS)~\cite{SDSS}; and finally from the
recent type Ia Supernova (SN) compilation of Ref.~\cite{Riess2004}.
We note that the cosmological parameters recovered from the data sets
used in this paper are in good agreement with subsequent studies made
by the SDSS collaboration using a different data set and a very
different theoretical modelling
(\cite{mcdonald2,seljaketal04,vwh,vh,vielandjulien05}). This demonstrates
that the
analysis of the Lyman-$\alpha$ forest QSO spectra is robust and that many
systematic uncertainties involved in the measurement are now better
understood than a few years ago.

The plan of the paper is as follows.
In section II we describe the notations we used for the isocurvature
sector. In section III we introduce the Lyman-$\alpha$ data that we
are employing. In section IV we discuss the general bounds on our full
AD+CDI parameter space from Lyman-$\alpha$, CMB, LSS and SN data using
a Bayesian likelihood analysis. We also check explicitly with a
hydrodynamical simulation the robustness of our Lyman-$\alpha$
data-fitting procedure, and we address the subtle issue of the role of
parametrizations and priors on the isocurvature bounds and in the
interpretations of our results.  We also discuss the specific curvaton
models with maximal anticorrelation and equal tilts for both adiabatic and
isocurvature modes, as well as bounds on double inflation models.
In section V we draw our conclusions.

\section{Mixed adiabatic/isocurvature models}

\subsection{Primordial spectra}

For the theoretical analysis, we will use the notation and some of the
approximations of Ref.~\cite{Beltran:2004uv}. During
inflation, more than one scalar field could evolve sufficiently slowly
that their quantum fluctuations perturbed the metric on scales larger
than the Hubble scale during inflation. These perturbations will later
give rise to one adiabatic mode and several isocurvature modes. We
will restrict ourselves here to the situation where there are only two
fields, $\phi_1$ and $\phi_2$, and thus only one isocurvature and one
adiabatic mode. Introducing more fields would complicate the
inflationary model and even then, it would be rather unlikely that
more than one isocurvature mode contributes to the observed
cosmological perturbations.  


Therefore, the two-point correlation function or power spectra of both
adiabatic and isocurvature perturbations, as well as their
cross-correlation, can be parametrized with three power laws, i.e.
three amplitudes and three spectral indices,
\begin{eqnarray}\label{primordial_P}\nonumber
\Delta_\calR^2(k)\!&\equiv&\!
{k^3\over2\pi^2}\langle\calR_{\rm rad}^2\rangle = {k_0^3\over2\pi^2}
A^2\left({k\over k_0}\right)^{\nad-1}\,,\\[2mm]
\Delta_\calS^2(k)\!&\equiv&\!
{k^3\over2\pi^2}\langle\calS_{\rm rad}^2\rangle = {k_0^3\over2\pi^2}
B^2\left({k\over k_0}\right)^{\niso-1}\,,\\[2mm]
\Delta_{\calR\calS}^2(k)\!&\equiv&\!{k^3\over2\pi^2}
\nonumber
\langle\calR_{\rm rad}\calS_{\rm rad}\rangle \, \nonumber\\[2mm]
&=& {k_0^3\over2\pi^2} A\,B\,\cos\Delta_{k_0} 
\left({k\over k_0}\right)^{\ncor+\half(\nad+\niso)-1}\,.\nonumber
\end{eqnarray}
Here, $\calR$ stands for the curvature perturbation, and
$\calS=(\delta_{\rm cdm}-3 \delta_{\gamma} / 4)$ for the CDI
perturbation. Both are evaluated during radiation domination and on
super-Hubble scales. We also introduced an arbitrary pivot scale $k_0$,
at which the amplitude parameters are defined through 
$A=\langle\calR_{\rm rad}^2\rangle^{1/2}$ and $B=\langle\calS_{\rm rad}^2\rangle^{1/2}$. In addition to the
fact that curvature and entropy perturbations are generally correlated
at the end of inflation, some extra correlation can be generated later by the
partial conversion of isocurvature into adiabatic perturbations.
The correlation angle $\Delta(k)$ is in general a function of $k$, and in
the above definitions, we approximated $\cos\Delta(k)$ by a power law
with amplitude $\cos\Delta_{k_0}$ and tilt $\ncor$. So, we assumed
implicitly that the inequality
\begin{equation}
\left|\cos\Delta_{k_0}\right| \left({k\over k_0}\right)^{\ncor} \leq 1
\end{equation}
holds over all relevant scales. We will enforce this condition in 
the following analysis.

\subsection{CMB anisotropy power spectra}


The angular power spectrum of temperature and polarization anisotropies
seen in the CMB today can be obtained from the radiation transfer
functions for adiabatic and isocurvature perturbations, $\Theta_l^{\rm
ad}(k)$ and $\Theta_l^{\rm iso}(k)$, computed from the initial conditions
$(\Rrad(k),\Srad(k)) =(1,0)$ and $(0,1)$, respectively, and convolved with the
initial power spectra,
\begin{eqnarray}\nonumber
C_l^{\rm ad}\!&\equiv&\,\int {dk\over k}\,
\left[\Theta_l^{\rm ad}(k)\right]^2\,
\left({k\over k_0}\right)^{\nad-1}\,,\\[2mm]
C_l^{\rm iso}\!&\equiv&\,\int {dk\over k}\,
\left[\Theta_l^{\rm iso}(k)\right]^2\,
\left({k\over k_0}\right)^{\niso-1}\,,\nonumber\\[2mm]
C_l^{\rm cor}\!&\equiv&\,\int {dk\over k}\,
\Theta_l^{\rm ad}(k)\,\Theta_l^{\rm iso}(k)\,
\left({k\over k_0}\right)^{\ncor+\half(\nad+\niso)-1}\,,\nonumber
\end{eqnarray}
Then, the total angular power spectrum reads
\begin{equation}
C_l = A^2 \, C_l^{\rm ad} + B^2\,C_l^{\rm iso} + 
2\,A\,B\,\cos\Delta_{k_0}\,C_l^{\rm cor}\,.
\end{equation}
In many works (see for instance~\cite{Amendola:2001ni,Peiris}),
the following parametrization is employed:
\begin{equation}
C_l = A^2 [ C_l^{\rm ad} + f_{\rm iso}^2\,C_l^{\rm iso} +
2 f_{\rm iso} \,\cos\Delta_{k_0}\,C_l^{\rm cor}]\,,
\end{equation}
where 
$f_{\rm iso} = B/A$ represents 
the entropy to curvature perturbation ratio during 
the radiation era at $k=k_0$. 
We will use here 
a slightly different notation,
used before by other groups~\cite{Crotty:2003aa,Langlois:1999dw,Stompor:1995py}:
\begin{equation}\label{alphanotation}
C_l = (A^2+B^2)\left[(1-\alpha)\,C_l^{\rm ad} + \alpha\,C_l^{\rm iso} + 2\beta\,
\sqrt{\alpha(1-\alpha)}\,C_l^{\rm cor}\right]\,,
\end{equation}
where $\alpha=B^2/(A^2+B^2)$ represents the isocurvature fraction at
$k_0$, and runs from purely adiabatic ($\alpha=0$) to purely
isocurvature ($\alpha=1$), while $\beta$ defines the correlation
coefficient at $k_0$, with $\beta=+1(-1)$ corresponding to maximally
correlated(anticorrelated) modes. There is an obvious relation between
both parametrizations:
\begin{equation}
\label{correspondance}
\alpha = f_{\rm iso}^2/(1+f_{\rm iso}^2)\,, \hspace{1cm} \beta = \cos\Delta_{k_0}\,.
\end{equation}
This notation has the advantage that the full parameter space of
$(\alpha,\ 2\beta\sqrt{\alpha(1-\alpha)})$ is contained within an
ellipse. The North and South rims correspond to fully
correlated ($\beta=+1$) and fully anticorrelated ($\beta=-1$)
perturbations, with the equator corresponding to uncorrelated
perturbations ($\beta=0$).  The East and West correspond to purely
isocurvature and purely adiabatic perturbations, respectively. Any
other point within the ellipse is an arbitrary admixture of adiabatic
and isocurvature modes.

We should emphasize that the three amplitude parameters $(A^2+B^2)$,
$\alpha$ and $\beta$ are defined at $k = k_0$, and that comparing
bounds from various papers is straightforward only when the pivot scale
is the same.
For instance, in the simple case where $\nad = \niso$, 
$\alpha$ is independent of $k_0$, 
but this is not the case for $\beta$:
if $\ncor > 0$, points within the
$(\alpha,\ 2\beta\sqrt{\alpha(1-\alpha)})$ ellipse are shifted
vertically toward the edges of the ellipse when one increases $k_0$
and shifted toward the horizontal $\beta = 0$ line when one
decreases $k_0$. When $\nad \neq \niso$, both $\alpha$
and $\beta$ depend on the pivot scale.
In addition, by changing the prior on the amplitudes,
a shift in the pivot scale affects 
the $\niso$ likelihood quite dramatically~\cite{Kurki-Suonio:2004mn}.
Throughout this paper, we will use $k_0 =
0.05$~Mpc$^{-1}$, which is the most frequent choice in the
literature. This value corresponds roughly to the multipole number
$l_0 \sim 400$. Therefore, the ratio $C_{400}^{\rm iso} / C_{400}^{\rm
ad}$ is roughly independent of the tilt values. For cosmological
parameters close to the best-fit $\Lambda$CDM model, one finds
$C_{400}^{\rm iso} / C_{400}^{\rm ad} \sim 0.01$. The
smallness of this number comes from the fact that 
$\Theta_l^{\rm iso}(k)$ is strongly suppressed with respect to
$\Theta_l^{\rm ad}(k)$ for large wavenumbers. Indeed,
the metric perturbations induced by isocurvature perturbations remain
small during radiation domination: so, for small scales entering early
inside the Hubble radius, the amplitude of the photon acoustic
oscillations is also small (as can be seen via its transfer function).  
As a consequence, even if during
radiation domination one has $\Srad(k_0) \sim \Rrad(k_0)$
(corresponding to $f_{\rm iso} \sim 1$ or $\alpha \sim 0.5$) the
isocurvature mode contributes only to 1\% of the observed anisotropy
near $l_0$. Of course, if $\niso$ is very different from
$\nad$, there could still be a large isocurvature contribution at
either larger or smaller scales.

\subsection{Matter power spectrum}
\label{mps}

Since in the following we will focus on the constraints induced on mixed
AD+CDI models
by the Lyman-$\alpha$ data, let us give a few details on the shape of the
linear matter power spectrum
\begin{eqnarray}
P(k) =  (A^2+B^2) &[(1-\alpha) \, P^{\rm ad}(k)+\alpha \, P^{\rm iso}(k)&
\nonumber\\
&+2\beta\sqrt{\alpha(1-\alpha)} \, P^{\rm cor}(k)]\,.&
\end{eqnarray}
Here $P^{\rm ad}$ and $P^{\rm iso}$ are computed from the initial
conditions $(\Rrad(k),\Srad(k)) =(1,0)$ and $(0,1)$ respectively,
exactly like
$\Theta_l^{\rm ad}(k)$ and $\Theta_l^{\rm iso}(k)$, 
and the cross-correlated term is
simply given by
\begin{equation}
P^{\rm cor}(k) = - (k/k_0)^{\ncor} [ P^{\rm ad}(k) P^{\rm iso}(k) ]^{\half},
\end{equation}
where the minus sign comes from the fact that with our definition of
${\cal S}$, a positive correlation $\langle \Rrad \Srad
\rangle >0$ in the early Universe
implies a reduction of the matter power spectrum today, and vice-versa.

In the limit $k \gg k_{eq}$, where $k_{eq}$ corresponds to modes
crossing the Hubble length at the time of equality, it is well known
(see e.g.~\cite{PK_analytic}) 
that the power spectra obey, to first approximation,
\begin{eqnarray}
P^{\rm ad}(k) &\propto& (k/k_0)^{\nad-4} \ln(k/k_{eq})^2 \,, \\
P^{\rm iso}(k) &\propto& (k/k_0)^{\niso-4} \,,
\label{shape_pk_iso}
\end{eqnarray}
which shows that for $\niso \simeq \nad$ the isocurvature contribution
to the small-scale power spectrum is generically much redder than the
adiabatic one.  The relative amplitude depends on the cosmological
parameters.  In the vicinity of the concordance $\Lambda$CDM model,
one finds $P^{\rm iso}(k_0)/P^{\rm ad}(k_0) \sim 4 \times 10^{-3}$ for
CDI. So, like for CMB anisotropies, we see that even when 
$\Srad(k_0) \sim \Rrad(k_0)$ in the early universe
(i.e. $f_{\rm iso}\sim 1$ or $\alpha\sim 0.5$), the isocurvature
contribution to the currently observed power spectrum is only of the 
per cent order, at least near the pivot scale. However, for large 
$\niso$, the contribution may be large on small scales.

\begin{figure*}[]
\vspace{1cm}
\includegraphics[angle=-90,width=8.5cm]{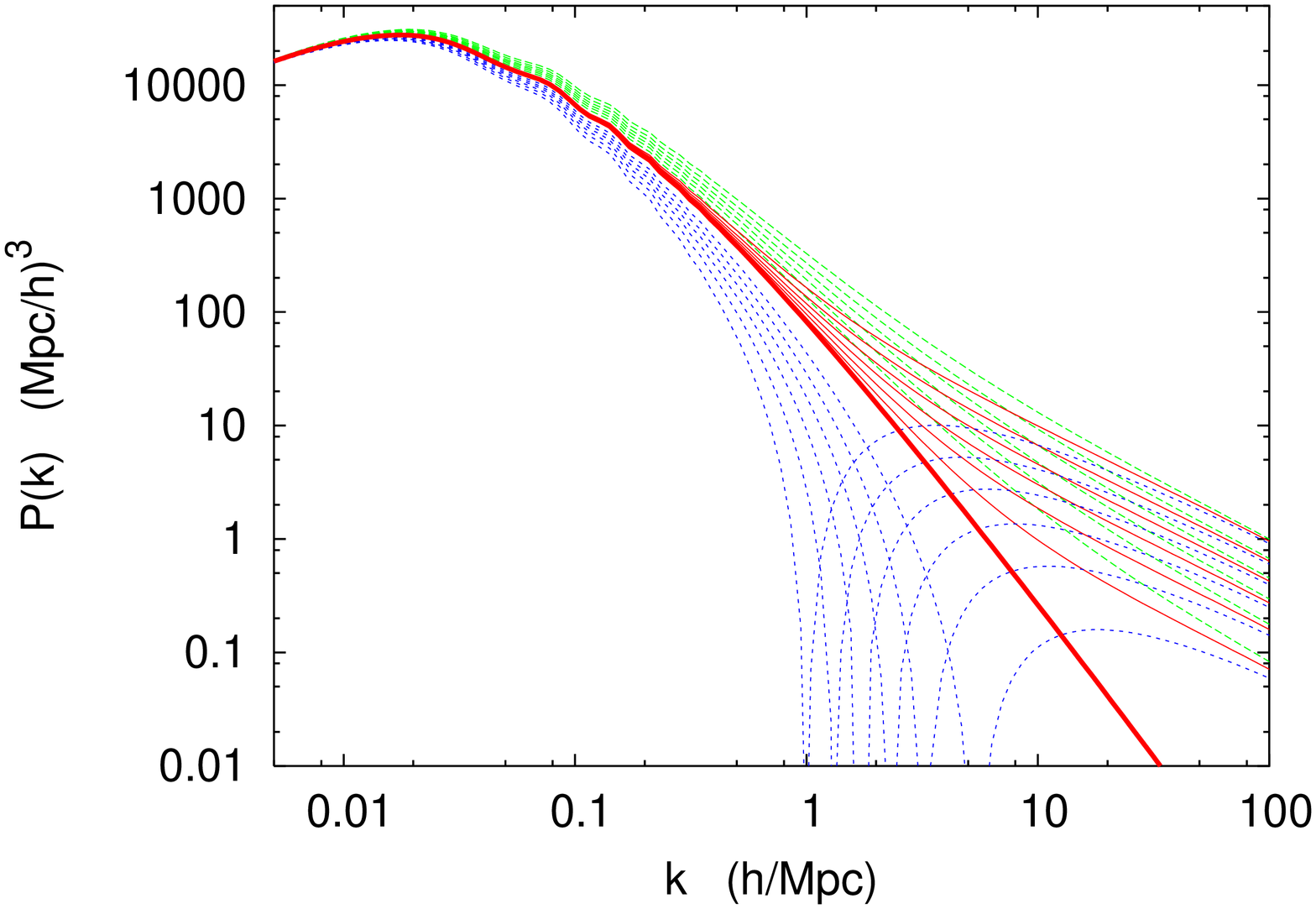}
\includegraphics[angle=-90,width=8.5cm]{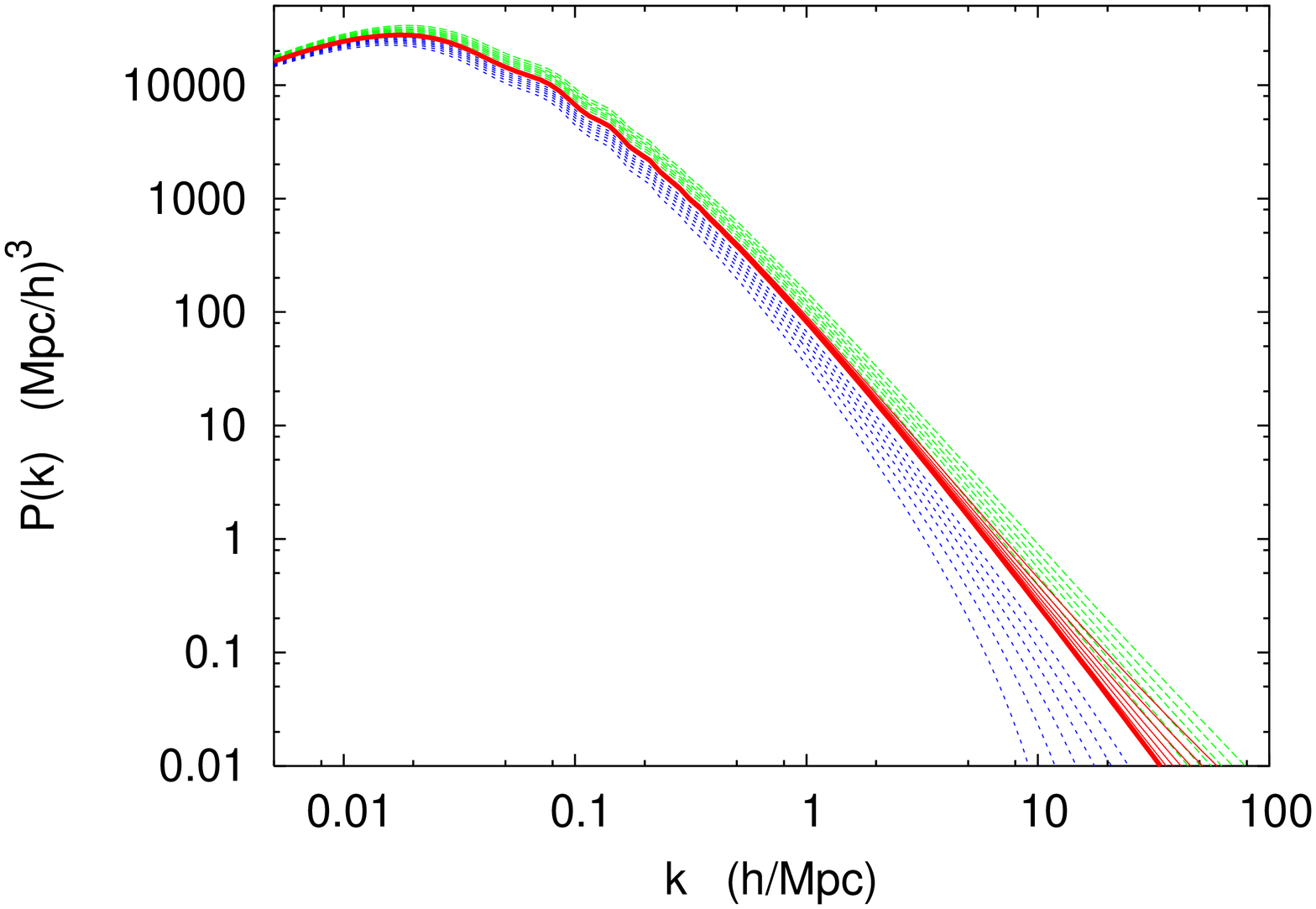}
\caption{\label{fig1} {\it(Left)} Matter power spectrum for a family
of mixed AD+CDI models, with all parameter fixed except
$\alpha$ and $\beta$ (the global normalization
also varies in order to mantain a fixed amplitude on large
scales). In particular, in all models we kept $\nad=0.95$, $\niso=3$
and $\ncor=0$. The thick line stands for the pure adiabatic case
($\alpha=0$). The thin solid (red) lines show uncorrelated models
($\beta=0$) with $\alpha=0.1,0.2,0.3,0.4,0.5,0.6$. The lower blue
(upper green) dashed lines show the maximally correlated models with
$\beta=1$ (anti-correlated with $\beta=-1$) for the same values
of $\alpha$.
{\it(Right)} Same as the left plot, but with $\niso$ reduced to $2.2$.}
\end{figure*}

Indeed, a large portion of the parameter region allowed by previous studies
corresponds to a significant isocurvature fraction $\alpha > 0.1$ and
to a very blue tilt $\niso > 1.5$. In this case, the matter power
spectrum is affected or even dominated by the non-adiabatic
contribution on small scales (typically for 
wavenumbers $k > 0.1 \, h/$Mpc).  We illustrate this
behavior on Fig.~\ref{fig1} for a particular set of AD+CDI models
with two different values of the isocurvature tilt, 
$\niso=3$ or $\niso=2.2$, and many possible values of $(\alpha, \beta)$.  
The impact of the non-adiabatic
contribution consists either in a smooth change of the effective slope
on small scales, or more radically in a sharp feature (a pronounced
break or a dip). The second situation can occur on relevant scales for
large positive $\beta$, and of course large enough values of $\alpha$
and $\niso$.

\section{Probing the matter power spectrum with the Lyman-$\alpha$
forest in QSO absorption spectra}

It is well established by analytical calculation and hydrodynamical 
simulations that the \lya forest blueward of the \lya emission line 
in QSO spectra  is produced by the inhomogeneous distribution of a
warm  ($\sim 10^4$ K) and photoionized intergalactic medium (IGM)
along the line of sight.  The opacity  fluctuations in the spectra 
arise from fluctuations in the matter density and trace the
gravitational clustering of the matter distribution in the
quasi-linear regime \cite{bi}. The \lya forest has thus been used 
extensively  as a probe of the matter power spectrum on comoving scales 
of  $(1-40) \, h^{-1}$Mpc \cite{bi,croft,vhs,mcdonald2}.

The \lya optical depth in  velocity space $u$ (km/s) is 
related to the neutral hydrogen distribution in real space 
as (see e.g. Ref. \cite{hgz}):
\begin{equation}
\tau(u)={\sigma_{0,\alpha} ~c\over H(z)} \int_{-\infty}^{\infty}
dy\, n_{\rm HI}(y) ~{\cal V}\left[u-y-v_{\parallel}(y),\,b(y)\right]dy \;,
\label{eqtau2} 
\end{equation} 
where $\sigma_{0,\alpha} = 4.45 \times
10^{-18}$ cm$^2$ is the hydrogen Ly$\alpha$ cross-section, $y$ is the real-space coordinate
(in km s$^{-1}$), ${\cal V}$ is the standard Voigt profile normalized
in real-space, $b=(2k_BT/mc^2)^{1/2}$ is the velocity dispersion in
units of $c$, $H(z)$ the Hubble parameter,
$n_{\rm HI}$ is the local density of neutral hydrogen and
$v_{\parallel}$ is the peculiar velocity along the line-of-sight. The
density of neutral hydrogen can be obtained by solving the
photoionization equilibrium equation (see e.g. \cite{katz}). 
The neutral hydrogen in the IGM responsible for the \lya forest
absorptions is highly ionized due to the metagalactic
ultraviolet (UV) background radiation produced by stars and QSOs at
high redshift. This optically thin gas in photoionization equilibrium
produces a \lya optical depth of order unity.

The balance between the photoionization heating by the UV background and adiabatic 
cooling by the expansion of the universe drives most of the
gas with $\delta_b< 10$, which dominates the \lya opacity, 
onto a power-law density relation $T=T_0\,(1+\delta_b)^{\gamma-1}$, where
the parameters $T_0$ and $\gamma$
depend on the reionization history and spectral shape of the UV
background and $\delta_b$ is the local gas overdensity ($1+\delta_b=\rho_b / \bar{\rho}_b$).

The relevant physical processes can be readily modelled in hydrodynamical 
simulations.  The physics of a photoionized IGM that traces the dark
matter distribution is, however, sufficiently simple that 
considerable insight can be gained from analytical 
modeling of the IGM opacity  based on the so called Fluctuating 
Gunn Peterson Approximation neglecting the effect of peculiar
velocities and the thermal broadening \cite{fgpa}. 
The Fluctuating  Gunn Peterson Approximation makes use of the
power-law temperature density relation and describes 
the relation between \lya opacity and gas density (see \cite{rauch,croft})
along a given line of sight as follows,
\begin{eqnarray}
&&\tau(z) \propto (1+\delta_b(z))^2 \, T^{-0.7}(z) 
= {\cal A}(z) \, (1+\delta_{b}(z))^{\beta} \, , 
\label{eqtau} \\
&&{\cal A}(z) = 0.433
\left(\frac{1+z}{3.5}\right)^6 
\left(\frac{\Omega_{b} h^2}{0.02}\right)^2
\left(\frac{T_0}{6000\;{\rm K}}\right)^{-0.7} \times \nonumber \\
&& \hspace{6mm} \left(\frac{h}{0.65}\right)^{-1}
\left(\frac{H(z)/H_0}{3.68}\right)^{-1} 
\left(\frac{\Gamma_{\rm HI}}
{1.5\times 10^{-12}\;{\rm s}^{-1}}\right)^{-1}\, ,\nonumber
\end{eqnarray}
where $\beta \equiv 2 - 0.7\, (\gamma-1)$ in the range $1.6-1.8$,
$\Gamma_{\rm HI}$  the HI photoionization rate, $H_0=h\, 100$ km/s/Mpc 
the Hubble parameter at redshift zero. 
For a quantitative analysis, however, full hydrodynamical simulations,  
which  properly simulate  the non-linear evolution of the
IGM and its thermal state, are needed.

Equations (\ref{eqtau2}) and (\ref{eqtau}) show how the  observed flux
$F=\exp{(-\tau)}$ depends on the underlying local gas density $\rho_b$, 
which in turn is simply related to the dark matter density, at least 
at large scales where the baryonic pressure can be neglected \cite{gh}. Statistical
properties of the flux distribution, such as the flux power spectrum,
are thus closely related to the statistical properties of the underlying
matter density field.

\subsection{The data: from the quasar spectra to the flux power spectrum}

The power spectrum of the observed flux in high-resolution \lya forest
data provides meaningful constraints on the dark matter power spectrum
on scales of $0.003\,{\rm s/km} < k < 0.03\,{\rm s/km}$, roughly
corresponding to scales of $(1-40) \, h^{-1}$Mpc (somewhat dependent on
the cosmological model). At larger scales the errors due to
uncertainties in fitting a continuum (i.e. in removing the long
wavelength dependence of the spectrum emitted by each QSO) become
very large while at smaller scales the contribution of metal
absorption systems becomes dominant (see e.g. \cite{kim04,mcdonald}).
In this paper, we will use the dark matter power spectrum that Viel,
Haehnelt \& Springel~\cite{vhs} (VHS) inferred from the flux power
spectra of the Croft et al.~\cite{croft} (C02) sample and the LUQAS
sample of high-resolution \lya forest data \cite{luqas}.  The C02
sample consists of 30 Keck high resolution HIRES spectra and 23 Keck
low resolution LRIS spectra and has a median redshift of $z=2.72$.
The LUQAS sample contains 27 spectra taken with the UVES spectrograph
and has a median redshift of $z=2.125$.  The resolution of the spectra
is 6 km/s, 8 km/s and 130 km/s for the UVES, HIRES and LRIS spectra,
respectively.  The S/N per resolution element is typically
30-50. Damped and sub-damped \lya systems have been removed from the
LUQAS sample and their impact on the flux power spectrum has been
quantified by \cite{croft}.  Estimates for the errors introduced by
continuum fitting, the presence of metal lines in the forest region
and strong absorptions systems have also been 
made~\cite{mcdonald,croft,hui,kim04}.

\subsection{From the flux power spectrum to the linear
matter power spectrum}
\label{hydro}

VHS have used numerical simulation to calibrate the relation between 
flux power spectrum and linear dark matter power spectrum  with
a method proposed by C02 and improved by \cite{gnedham}
and VHS.  A set  of hydrodynamical simulations for a coarse
grid of the relevant parameters is used  to find a model that  provides
a reasonable  but not exact fit to the observed flux power
spectrum. Then, it is assumed that the differences
between the model and the observed linear power spectrum depend linearly 
on the matter power spectrum.  

The hydrodynamical simulations are used to determine a bias function
between flux and matter power spectrum: $ P_F(k) = b_F^2(k)\;P(k)$, on
the range of scales of interest. In this way the linear matter power 
spectrum can be recovered with reasonable computational 
resources.\footnote{Note that this bias is different from the usual 
bias between light and matter, and can be strongly scale-dependent.} 
This method has been found to be robust provided the systematic 
uncertainties are properly taken into account \cite{vhs,gnedham}.
Running hydrodynamical simulations for a fine grid of all the relevant
parameters is unfortunately computationally prohibitive (see discussion
in \cite{vh} on a possible attempt to overcome this problem).

We have seen in section \ref{mps} that the isocurvature mode
contribution can create distortions in the small-scale linear matter
power spectrum.  Of course, this extra freedom was not taken into
account in the definition of the grid of models in VHS. In principle,
we should run simulations for a new grid with extra parameters
($\alpha$, $\beta$, $\niso$, $\ncor$). Alternatively, we can carry a
tentative analysis with the same function $b_F(k)$ and the same error
bars as in the pure adiabatic case, and check the validity of our
results {\it a posteriori}. The idea is simply to select a
marginally excluded model with the largest possible deviation from
adiabaticity in the matter power spectrum. For this model, we 
run a new hydrodynamical simulation and we
compare $P_F(k)/P(k)$ with the function $b_F^2(k)$ used in the
analysis. In case of good agreement, the results will be validated.
We expect this agreement to be fairly good on large scales, but 
deviations should appear on small scales, because of the different non-linear
evolution.

The  use of state-of-the-art hydrodynamical simulations is a significant 
improvement  compared to previous studies which used 
numerical simulation of dark matter only \cite{croft}.
We use  the parallel TreeSPH code GADGET-II
\cite{volker} in its TreePM mode which speeds up the
calculation of long-range gravitational forces considerably. The
simulations are performed with periodic boundary conditions with an
equal number of dark matter and gas particles. Radiative cooling and
heating processes are followed using an implementation similar to
\cite{katz} for a primordial mix of hydrogen and helium. The UV
background is given by \cite{haardt}. To maximise the speed of
the simulation a simplified criterion of star formation has been
applied: all the gas at overdensities larger than 1000 times the mean
overdensity is turned into stars \cite{vhs}.
The simulations were run on {\sc cosmos}, a 152 Gb shared memory Altix 3700
with 152 CPUs hosted at the Department of Applied Mathematics and
Theoretical Physics (Cambridge).

\subsection{Systematics Errors}

There is a number of systematic uncertainties and statistical errors
which affect the inferred power spectrum and an extensive discussion
can be found in \cite{croft,gnedham,vhs,vh}. VHS estimated the 
uncertainty of the overall {\it rms} fluctuation amplitude of matter 
fluctuation to be 14.5 \% with a wide range of different factors 
contributing.  

We present here a brief summary. The effective optical
depth, $\tau_{\rm eff}=-\ln\langle F\rangle$ which is essential for the
calibration procedure has to be determined separately from the
absorption spectra. As discussed in VHS, there is a considerable
spread in the measurement of the effective optical depth in the
literature.  Determinations from low-resolution low S/N spectra
give systematically higher values than high-resolution  high S/N
spectra. However, there is little doubt that the lower values from
high-resolution high S/N spectra are appropriate and the range
suggested in VHS leads to a 8\% uncertainty in the {\it rms}
fluctuation amplitude of the matter density field (see Table 5 in VHS).
Other uncertainties are the slope and normalization of the
temperature-density relation of the absorbing gas which is usually
parametrised as $T=T_0\,(1+\delta_b)^{\gamma-1}$.  $T_0$ and $\gamma$
together contribute up to 5\% to the error of the inferred fluctuation
amplitude. VHS further estimated that uncertainties due to the C02
method (due to fitting the observed flux power spectrum with a bias
function which is extracted at a slightly different redshift than the
observations) contribute  about 5\%. They further
assigned a 5 \% uncertainty to the somewhat uncertain effect of
galactic  winds and finally an   8\%  uncertainty due the
numerical simulations (codes used by different groups give somewhat
different results).  Summed in quadrature, all
these errors led to the estimate of the overall uncertainty of 14.5\%
in the {\it rms} fluctuation amplitude of the matter density field.

For our analysis we use the inferred DM power spectrum in the range
$0.003\,{\rm s/km} < k < 0.03\,{\rm s/km}$ as given in Table 4 of VHS.
(Note that, as in ~\cite{vielandjulien05} we have reduced the power
spectrum values by 7\% to mimick a temperature-density relation with $\gamma=1.3$, the middle
of the plausible range for $\gamma$ \cite{temperature}).

Unfortunately at smaller scales the systematic
errors become prohibitively large mainly due to the large contribution
of metal absorption lines to the flux power spectrum (see Fig. 3 of
Ref.~\cite{vhs}) and due to the much larger sensitivity of the flux power
spectrum to the thermal state of the gas at these scales.

\begin{figure}[]
\includegraphics[angle=0,height=9cm]{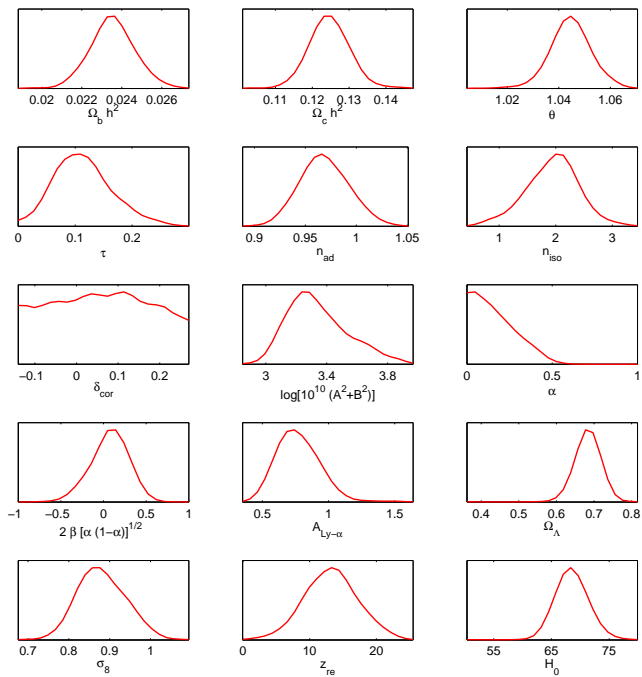}
\caption{\label{fig_lya_cdi} Likelihood for
the AD+CDI model, using all our data set. The first eleven parameters
are independent, while the last four are related parameters
(with non-flat priors).
}
\end{figure}
\begin{table}[]
\begin{center}
\begin{tabular}{c|c}
parameter & $1\sigma$ C.L. \\
\hline
\hline
$\Omega_b h^2$ & 0.0235 $\pm$ 0.0011 \\
$\Omega_c h^2$ & 0.125 $\pm$ 0.005\\
$\theta$ & 1.045 $\pm$ 0.008 \\
$\tau$ & 0.11 $\pm$ 0.05\\
$n_{\rm ad}$ & 0.97 $\pm$ 0.02 \\
$n_{\rm iso}$ & 1.9 $\pm$ 0.5 \\
$\delta_{\rm cor}$ & {\rm within~prior~range} \\
$\log[10^{10} (A^2+B^2)]$ & 3.3 $\pm$ 0.2 \\
$\alpha$ & $<0.20$ \\
$2 \beta [\alpha(1-\alpha)]^{1/2}$ & 0.1 $\pm$ 0.2 \\
$A_{{\rm Ly}-\alpha}$ & 0.8 $\pm$ 0.2 \\
\hline
$\Omega_{\Lambda}$ & 0.68 $\pm$ 0.03 \\
$\sigma_8$ & 0.88 $\pm$ 0.06 \\
$z_{\rm re}$ & 13 $\pm$ 4 \\
$H_0$ & 69 $\pm$ 3 \\
\end{tabular}
\end{center}
\caption{\label{tab_lya_cdi}
$1\sigma$ confidence limits for the AD+CDI model, using all our data set,
for the eleven basis parameters with flat priors,
and below, for related parameters.
}
\end{table}
\begin{figure}[]
\includegraphics[angle=0,height=8cm]{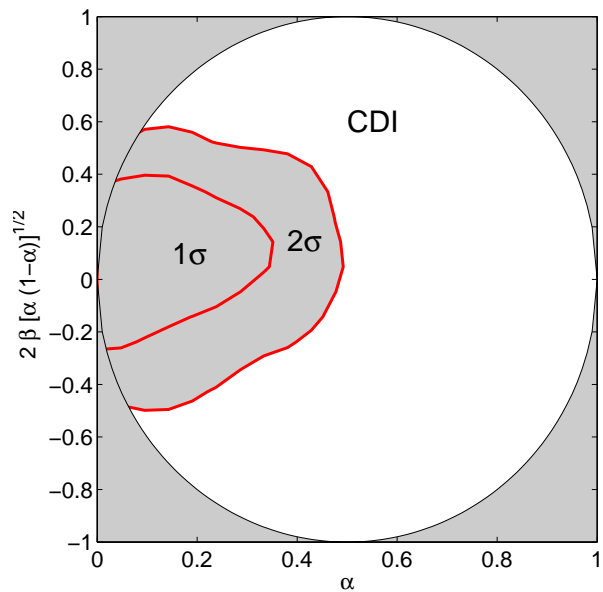}
\caption{\label{fig_lya_cdi_2D} Two-dimensional likelihood for
the amplitude of the isocurvature mode and of the cross-correlation component,
near the pivot scale. We adopted a flat prior within the ellipse 
(which appears here as a circle) in which these
parameters are defined.
}
\end{figure}

\section{Fitting the data}

\subsection{Parameter basis and priors}

Any AD+CDI model is described by the usual six parameters of the
$\Lambda$CDM model, plus four parameters for the isocurvature sector
(two amplitudes and two tilts). Like in most of the literature, we
define the amplitudes parameters at the pivot scale
$k_0=0.05\,$Mpc$^{-1}$. For the isocurvature fraction, we could decide
to impose a flat prior on $\fiso$, or $\alpha$, or any function of
them; different choices are not equivalent, in general.  We will come
back to the dependence of the final result on the choice of priors in
section~\ref{sec:priors}. Meanwhile, we chose a specific set of
parameters which appear linearly in the expression of the observable
power spectra, $\alpha$ and $2\beta \sqrt{\alpha(1-\alpha)}$, and that
we believe are physically relevant. As already mentioned, these two
parameter are defined within an ellipse, in which we assume a flat
prior.  Furthermore, we must take into account the inequality
\begin{equation}
\label{ncorrconst}
\left|\cos\Delta \right| =
\left|\beta \right| \left({k\over k_0}\right)^{\ncor} \leq 1
\end{equation}
which should hold at least over the scales probed by the data, i.e.
typically between $k_{\rm min} = 4 \times 10^{-5}$Mpc$^{-1}$ and
$k_{\rm max} = 2\,$Mpc$^{-1}$. This is achieved
by introducing a
new parameter
$\delta_{\rm cor} \equiv n_{\rm cor} / \ln |\beta|^{-1}$, with a flat prior within
the range $-0.14 < \delta_{\rm cor} < 0.27$.
In summary, our basis parameters with flat priors consists of:
\begin{itemize}
\item the baryon density, $\oB=\OB h^2$,
\item the cold dark matter density, $\omega_c = \Omega_c h^2$,
\item the ratio $\theta$ of the sound horizon to the
angular diameter distance multiplied by 100,
\item the optical depth to reionization, $\tau$,
\item the adiabatic tilt, $\nad$,
\item the isocurvature tilt, $n_{\rm iso}$,
\item the parameter related to the
tilt of the cross-correlation angle, $\delta_{\rm cor} \in [-0.14,0.27]$,
\item the overall normalization, $\ln[10^{10} (A^2+B^2)]$,
\item the isocurvature fraction, $\alpha$,
\item the cross-correlation amplitude, $2\beta\sqrt{\alpha(1-\alpha)}$.
\end{itemize}
In addition, there are three independent parameters related to
observations: the Lyman-$\alpha$ calibration parameter $A_{{\rm Ly}-\alpha}$ 
defined in \cite{vwh}, on which we impose the same Gaussian prior
$A_{{\rm Ly}-\alpha}=1.0\pm 0.29$; and the two bias
parameters associated to the 2dF and SDSS data with flat priors.
Our full parameter space is therefore 13-dimensional.

\subsection{Results}
\label{sec:results}

We compute the marginalized Bayesian likelihood of each
parameter with a Monte Carlo Markov Chain method, using the public
code CosmoMC~\cite{cosmomc}. The results are displayed in
Fig.~\ref{fig_lya_cdi} and Table~\ref{tab_lya_cdi}
(after marginalization over the 2dF and SDSS bias
parameters).  The data favors purely adiabatic models, but remains
compatible with an isocurvature fraction $\alpha<0.40$ at the
2$\sigma$ (95\%) confidence level (CL), with a tilt $\niso=1.9 \pm
1.0$ ($2\sigma$ CL). The one-dimensional likelihoods for $\alpha$,
$2\beta\sqrt{\alpha(1-\alpha)}$ must be interpreted with
care: the fact that these parameters are defined 
within an ellipse implies that there
is more parameter space available near $\alpha=0.5$ and
$2\beta\sqrt{\alpha(1-\alpha)}=0$.  More interesting are
the two-dimensional likelihood contours for $(\alpha, \,
2\beta\sqrt{\alpha(1-\alpha)})$ displayed in
Fig.~\ref{fig_lya_cdi_2D}, since in
this representation the prior is really flat inside the
ellipse. From this figure, it is clear that the data prefers an
uncorrelated isocurvature contribution. The flatness of the
$\delta_{\rm cor}$ likelihood shows that the data give no
indication on the tilt of the cross-correlation angle. 

\begin{figure}[]
\includegraphics[angle=0,height=4.2cm]{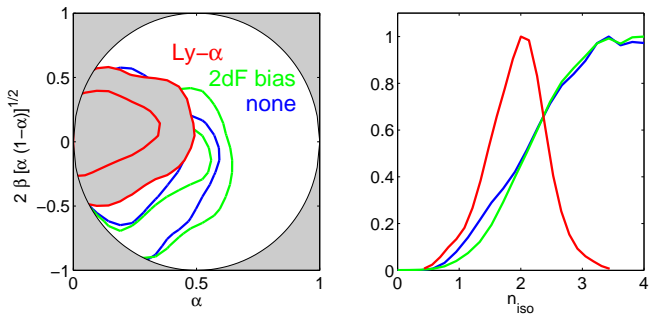}
\caption{\label{lya_vs_bias_cdi} Likelihood of the
isocurvature-related parameters, for the three
combinations of data sets described in section \ref{impact}:
``Lyman-$\alpha$'' (red), ``2dF bias prior'' (green) and ``none'' (blue).
{\it(Left)} Marginalized 1$\sigma$ and 2$\sigma$ confidence levels in
the $(\alpha, 2 \beta [\alpha (1- \alpha)]^{1/2})$
space. {\it(Right)} Marginalized probability distribution for $\niso$.}
\end{figure}
\begin{figure}[]
\includegraphics[angle=0,height=4.2cm]{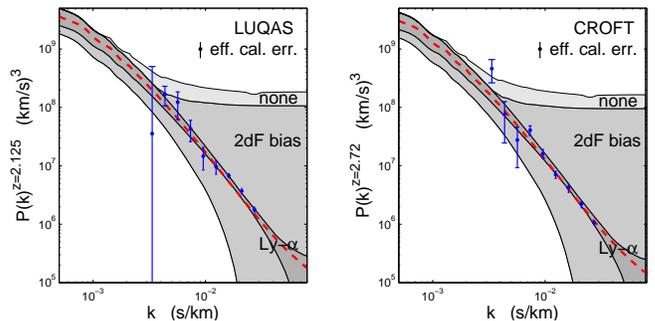}
\caption{\label{fig_lya_pk} Favored ranges for the matter power
spectrum $P(k)$ in the three runs ``Lyman-$\alpha$'' (dark), ``2dF
bias prior'' (medium) and ``none'' (light), compared with our
Lyman-$\alpha$ data, from the LUQAS quasar spectra {\it (left)} and
from the re-analyzed Croft et al. spectra {\it (right)}. The bands
represent the envelope of all the matter power spectra in the Markov
chains (after eliminating models with the worse likelihood). 
Each power spectrum has been computed at the median redshift
of the data and re-expressed in units of km/s. In
addition to the statistical errors, the data points share an overall
effective calibration error, whose standard deviation is displayed in
the top right corners. For the run including the Lyman-$\alpha$ data,
each power spectrum has been divided by the value of the calibration
parameter. The red dashed curves show the particular power spectrum
discussed in section~\ref{posteriori}.}
\end{figure}

\subsection{Specific impact of the Ly-$\alpha$ data}
\label{impact}

The Lyman-$\alpha$ forest provides a powerful indication on both the
amplitude and the shape of the matter power spectrum for
$k> 0.01$ s/km, i.e. roughly larger than $1h/$Mpc.
In order to illustrate the importance of this
data set in our results, we repeat the same analysis without
Lyman-$\alpha$ data.  In this case, there are two options: we can
either use the 2dF and SDSS galaxy power spectrum data as a constraint
only on the {\it shape} of the matter power spectrum, as already done
in the previous analysis of section~\ref{sec:results}; 
or introduce a bias prior derived e.g. from
the third and fourth-order galaxy correlation function of the 2dF
catalogue \cite{2dFGRS,Verde2003}, in order to keep an information on
the amplitude of the matter power spectrum\footnote{Technically, our
bias prior is implemented in the same way as in Ref.\cite{Cuoco}:
see Eq.~(27) and following lines in this reference.}.

For these three cases, that we call ``Lyman-$\alpha$'', ``2dFbias prior''
and ``none'', the 2$\sigma$ upper bound on $\alpha$ are respectively equal
to 0.4, 0.5 and 0.5. The likelihoods for the most interesting parameters are displayed in Fig.~\ref{lya_vs_bias_cdi}. As expected, the
Lyman-$\alpha$ data set is significanty more powerful than the 2dF bias prior
for cutting out models with large $\alpha$,
and even more clearly, with
large $\niso$ or large anticorrelation, as can be seen in Fig.~4.
It is important to note that without
these data, all results depend on our arbitrary prior $\niso<4$:
values far beyond this upper bound could still be compatible with the
data, as also found in Ref.~\cite{Kurki-Suonio:2004mn} when using the
same pivot scale. In the presence of the Lyman-$\alpha$ data, we get a
robust upper
bound on $\niso$, and none of our priors play a role in the final results,
with the exception of the well-motivated $\delta_{\rm cor}$ prior.

The impact of the Lyman-$\alpha$ data can be understood visually from
Fig.~\ref{fig_lya_pk}. After running each case, we consider the
collection of all matter power spectra in our Markhov chains (except
models with a bad posterior likelihood ${\cal L} < {\cal L}_{\rm
max}/5$).  The gray bands in Fig.~\ref{fig_lya_pk} correspond to the
envelope of all these $P(k)$'s, compared to the Lyman-$\alpha$ data
points. As expected, when the Lyman-$\alpha$ is not used, the band
gets very wide above the wavenumber $k \sim 0.2~ h/$Mpc$~ \sim 2
\times 10^{-3}\,$s/km (note that for models with $\niso=4$, the small-scale
power spectrum is asymptotically flat).
The role of the bias prior is
marginal: it simply favors models with the lowest global
normalization, but without affecting the isocurvature fraction and
tilt.  Using the Lyman-$\alpha$ data, we can exclude any break in the
power spectrum on scales $k \le 3 ~h/$Mpc$~\sim ~ 3 \times
10^{-2}\,$s/km.  This results in much stronger constraints for the
parameters $(\alpha, \beta, \niso)$, as can be seen from Fig.~4.

\begin{figure}[]
\vspace{-0.3cm}
\includegraphics[angle=0,height=9cm]{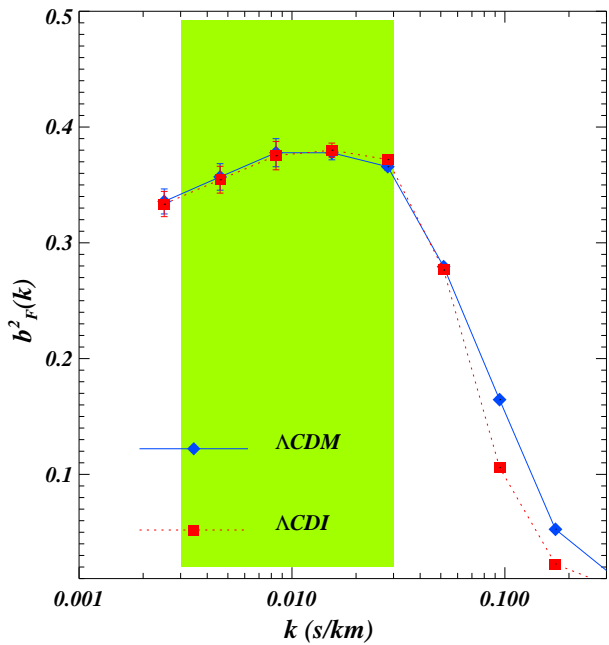}
\caption{\label{bias} The ratio 
$b_F(k)^2 \equiv P_{\rm flux}(k)/P_{\rm matter}(k)$ at $z=2.75$, 
computed from the hydrodynamical simulations as the ratio
of the flux power spectrum (averaged over 1000 line-of-sights)
over the input linear matter power spectrum. The solid blue curve shows the
result for an adiabatic $\Lambda$CDM close to the best-fit
model, while the dashed red curve was obtained from the ``most extreme
AD+CDI mixed model'' defined in section~\ref{posteriori} and here labelled as $\Lambda$CDI.
The green band shows the region in which the Lyman-$\alpha$ data
is used in the present analysis.}
\end{figure}

\subsection{Checking the validity of the Ly-$\alpha$ data fitting procedure}
\label{posteriori}

We apply the strategy described in section~\ref{hydro} in order to check
the validity of our Lyman-$\alpha$ data fitting procedure. We take the large
number of samples contained in our Markov chains, and
eliminate all models with a likelihood smaller than 
${\cal L}_{\rm max}/10$ (in terms of effective $\chi^2$, this corresponds to 
$\Delta \chi^2 = \chi^2 - \chi_{\rm min}^2 > 20$). 
We then select the model with the largest value of 
$\alpha$, which represents the strongest deviation from the purely adiabatic 
model. The corresponding matter power spectrum is plotted in 
Fig.~\ref{fig_lya_pk} and has a break around 
$k \sim 5 ~h/$Mpc$ ~ \sim 5 \times 10^{-2}\,$s/km. Above this wavenumber, 
the slope of the power spectrum is given
by eq.~(\ref{shape_pk_iso}) with $\niso=2.7$.
For this ``extreme'' model, we perform a hydrodynamical simulation
as described in section~\ref{hydro}, and compare the bias function
$b_F(k)$ with that assumed throughout the analysis.
As shown in Fig.\ref{bias}, in the range $0.003 < k < 0.3 \,$km/s
probed by the data, the difference
between the two functions is very small with respect to the statistical
errors on the data.
We conclude that in the present context, our Lyman-$\alpha$
data fitting procedure is robust, and does not
introduce an error in the 1$\sigma$ or 2$\sigma$ bounds
derived for each parameter of the AD+CDI mixed model.

\begin{figure}[]
\includegraphics[angle=0,height=4.2cm]{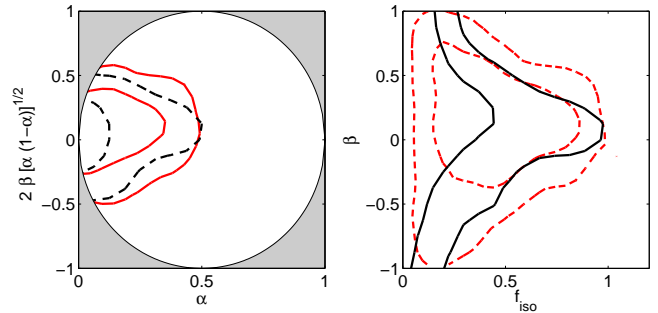}
\caption{\label{lya_cdi_op_vs_fiso} Two-dimensional 68\% and 95\%
confidence limits for the
CDI mode amplitude and cross-correlation angle (evaluated at the
pivot scale). The two plots show the results of two independent
runs with different parameter basis and priors. On the left
(solid red curves),
the parameters are $(\alpha, \, 2 \beta \sqrt{\alpha(1-\alpha)})$,
with a flat prior within the ellipse.
On the right (solid black lines), the parameters are
$(f_{\rm iso}, \cos\Delta)$, related to the previous parameters
through eqs.(\ref{correspondance}),
with a flat prior within the square.
The dashed curves show for comparison
the likelihood contours obtained for one parameter set, assuming
a flat prior on the {\it other} parameter set.
}
\end{figure}

\subsection{The role of parametrization and priors}
\label{sec:priors}

The fact of choosing a top-hat prior in the
$(\alpha, \, 2 \beta \sqrt{\alpha(1-\alpha)})$
parameter space is rather arbitrary. Other groups prefer
to take top-hat priors on $f_{\rm iso}$, defined in
Eq.(\ref{correspondance}), and $\cos\Delta=\beta$. Due to the
non-linear transformation between the two basis, they are clearly not
equivalent in terms of priors (see the discussion of this point
in \cite{Trotta:2005ar}, 
in the context of Bayesian Evidence calculation for
adiabatic versus mixed models).

We checked this issue explicitly with an independent run
based on the $(f_{\rm iso}, \cos\Delta)$ basis. The results are summarized
in Fig.~\ref{lya_cdi_op_vs_fiso}. As expected from the Jacobian,
the $(f_{\rm iso}, \cos\Delta)$ option gives more weight to models with
a small isocurvature fraction. For instance, the run with a flat prior
on $f_{\rm iso}$ gives a 1$\sigma$ bound $f_{\rm iso} < 0.26$, while that
with a flat prior on $\alpha$ gives $f_{\rm iso} < 0.66$. However,
at the 2$\sigma$ level, the relative difference is small ($f_{\rm iso} <0.75$
versus $f_{\rm iso} <0.87$) because the Jacobian is asymptotically flat.

In principle, in Bayesian analysis, the choice of parameter basis and
priors should reflect one's knowledge on the model before comparision
with the data. However, in the absence of a unique underlying physical
model motivating the presence of isocurvature modes, different
scientists might put foward different choices of prior. This intrinsic
freedom in Bayesian analyses should always be kept in mind when
quoting bounds, especially for parameters which represent physical
ingredients not strictly needed by the data, which is the case here
for the isocurvature sector parameters (for other parameters such that
the data picks up a narrow allowed region, a change of priors won't
affect the bounds very much). However, even for the isocurvature
parameters discussed here, it is reassuring to see from our analysis
that the 2$\sigma$ contours obtained from the the two runs and
compared in Fig.~\ref{lya_cdi_op_vs_fiso} are roughly in agreement.

\begin{figure}[]
\includegraphics[angle=0,height=4.cm]{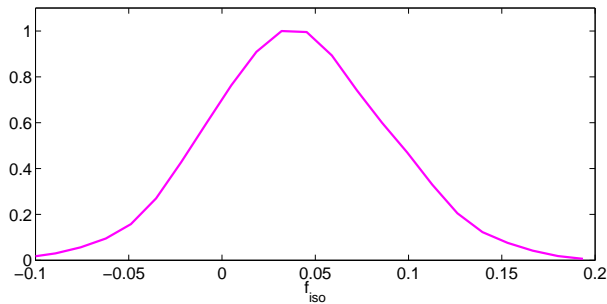}
\caption{\label{fig_curv} Likelihood for $f_{\rm iso}$ in the curvaton
model.
}
\end{figure}

\subsection{The curvaton model}

In this section we derive bounds on the specific case of curvaton models.
The curvaton hypothesis is an ingenuous way to generate the observed
curvature perturbation from a field (the curvaton) different from that
which drives inflation (the inflaton)~\cite{Lyth:2001nq}. In practice
there is not much difference in the phenomenological signatures left
in the CMB and LSS compared to an ordinary inflationary model. However,
there are a few cases in which it is
possible to leave a ``residual" isocurvature component, together with the
dominant curvature contribution. More specifically, in the curvaton
models in which the curvaton field is responsible for
the CDM component of matter, there are various possibilities depending
on the time of creation of CDM versus the decay of the curvaton field.
In all these cases, the curvature and isocurvature perturbations are
related to the gauge invariant Bardeen variable $\zeta$ as
\begin{eqnarray}
&&\calS=3(\zeta_{\rm cdm} - \zeta)\,,\\[2mm]
&&\calR=-\zeta\,.
\end{eqnarray}
Let us classify here the different cases: 1) when CDM-creation occurs before
the curvaton decays and the fraction $r$ of the total energy density
in the curvaton field at the time of its decay is negligible. Then
$\calS=-3\zeta=3\calR$, which corresponds to $f_{\rm iso} = 3$
($\alpha=0.9$), and $\beta=+1$ (maximally correlated), with $\niso=\nad$;
2) when CDM-creation occurs before the curvaton decays but the fraction $r$
at decay is important. This case requires specific model input and in
principle can have any value of $f_{\rm iso}$ and $\beta$, while
$\niso=\nad$; 3) when CDM-creation occurs at the decay of the curvaton
and the fraction $r<1$. In this case, $\zeta_{\rm cdm} = \zeta/r$
and thus $\calS = 3(r^{-1}-1)\zeta= 3(1-1/r)\calR$, which corresponds
to $f_{\rm iso} = 3(1-1/r)$, i.e. $\beta=-1$ (maximally anticorrelated)
and $\niso=\nad$; 4) when CDM-creation occurs after the curvaton decay.
Then there is only one thermal fluid in equilibrium, $\zeta_{\rm cdm} =
\zeta$, and there is no way to generate an isocurvature perturbation,
$\calS=0$.

Since case 1) is already excluded at many sigma,
 and case 2) is essentially identical (except for
$\niso=\nad$) to our generic analysis, we will concentrate on case 3)
of a maximally anticorrelated mixture of isocurvature and adiabatic modes
with equal tilts and $\delta_{\rm cor}=0$. Our results are summarized in
Fig.~\ref{fig_curv}, which shows the likelihood distribution for the
generic curvaton model. We have used $\nad=\niso$, $\delta_{\rm cor}=0$,
and $\beta=\pm1$,
which is equivalent to $\beta=1$ and $\fiso$ positive or negative:
$f_{\rm iso}>0$ corresponds to $\beta=1$, or positive
correlation between $\Rrad$ and $\Srad$, i.e. suppression of power in
$P(k)$ and in the large-scale CMB temperature spectrum; while $f_{\rm
iso}<0$ corresponds to the opposite anti-correlated case.

We find $f_{\rm iso} = 0.04 \pm 0.09$ at the 2$\sigma$-level, which
implies $r>0.98$ at the same CL. In our opinion, such a stringent
constraint on the fraction of energy density in the curvaton at decay
calls for a tremendous finetuning (there is no physical reason to expect
that the curvaton should decay precisely when it is starting to dominate
the total energy density of the universe, within 2\%), which makes 
the curvaton hypothesis in its most attractive scenario very unlikely.

\subsection{The double inflation model}

Another chance to generate an observable isocurvature signature 
is through the possible presence of two scalar fields driving
inflation \cite{Polarski:1994rz, double}.
The simplest case at hand is that of
two massive fields coupled only gravitationally:

\begin{equation}\label{double_lagrangian}
\mathcal L =\frac{1}{2}(\phi_{h;\mu}\phi_{h}^{;\mu}-m_h^2\phi^2_h)
+\frac{1}{2}(\phi_{l;\mu}\phi_{l}^{;\mu}-m_l^2\phi^2_l)~,
\end{equation}
where $m_h$ and $m_l$ are the masses of the heavy and light fields respectively.\\
  
We assume slow-roll conditions during inflation, and use the number of e-folds till 
the end of inflation 
$s=-\ln(a/a_{\rm end})$ to parametrize the fields as:

\begin{equation}\label{double_fields}
\phi_h^2=\frac{s}{2\pi G}\sin^2\theta;~\phi_l^2=\frac{s}{2\pi G}\cos^2\theta~,
\end{equation}

Using the field and Friedmann equations, we can solve 
for the rate of expansion during inflation:

\begin{equation}\label{hubble_double}
H^2(s)\simeq \frac{2}{3}s\cdot m_l^2\lbrack 1+(R^2-1)\sin^2\theta\rbrack~,
\end{equation}
where $R= m_h/m_l$, and find the number of e-folds as a function of $\theta$:

\begin{equation}
s(\theta)=s_0 \frac{(\sin\theta)^{2/(R^2-1)}}
{(\cos\theta)^{2R^2/(R^2-1)}}~.
\end{equation}

The perturbed Einstein equations can be solved for long wavelength 
modes in the longitudinal gauge. Assuming that the heavy 
field decays into CDM 
whereas the ligth field produces other species, we 
find the magnitudes of the curvature and entropy perturbation at 
horizon crossing. 
During radiation domination and for super-Hubble modes, this gives:

\begin{eqnarray}
\Rrad(k) &=& -\sqrt{\frac{4 \pi G}{k^3}} H_k
s_k^{1/2} \Big(\sin\theta_k\,e_h({\bf k}) + 
\cos\theta_k\,e_l({\bf k})\Big)\,\nonumber
\\
\Srad(k) &=& \sqrt{\frac{4 \pi G}{k^3}} H_k s_k^{-1/2}
\left({e_h({\bf k})\over \sin\theta_k} - {R^2\,e_l({\bf k})\over 
\cos\theta_k}\right)\,
\end{eqnarray}
where $e_i({\bf{k}})$ are gaussian random fields associated 
with the quantum fluctuations of the fields, and the subindex $k$ 
implies the value of the corresponding quantity at horizon crossing
during inflation. One typically expects $s_k\simeq 60$. 
It can be seen 
from (\ref{primordial_P}) that the correlation power spectrum 
has no scale dependence, and thus, for this model $\ncor=0$,
while the adiabatic and isocurvature tilts have expressions
\begin{eqnarray}
\nad &=& 1 -{2\over s_k} + {(R^2-1)\tan^2\theta\over 
2s_k(1+R^2\tan^2\theta)^2}\,,\\[2mm]
\niso &=& 1 - {(R^2-1)(R^2\tan^4\theta-1)(1+\tan^2\theta)\over 
s_k(1+R^2\tan^2\theta)^2(1+R^4\tan^2\theta)}\,,
\end{eqnarray}
whose values, for $s_k=60$, are typically $\nad=0.97$ and $\niso$
in the range $[0.97,0.90]$ for $R\in[1,4]$. Since $\niso > 0.93$
at 95\% c.l., models with large values of $R$ are ruled out.

It was shown in \cite{Beltran:2004uv} that a 
relationship beteween $\alpha$ and $\beta$ can be found. It can be 
simply expressed as a straight line in our parameter space: 
\begin{equation}\label{relation} 
2\beta\sqrt{\alpha(1-\alpha)}=\frac{2(R^2-1)}{s_k}(1-\alpha)\,.
\end{equation}
On the other hand, for these models, the parameters $\alpha$ and $\beta$
have minimum and maximum values respectively, which only depend on 
the ratio $R$ and the number of e-folds $s_k$,
\begin{eqnarray}
\alpha_{\rm min} &=& {(R^2+1)^2\over s_k^2+(R^2+1)^2}\,,\\[2mm]
\beta_{\rm max} &=& {R^2-1\over R^2+1}\,,\\[2mm]
\left.2\beta\sqrt{\alpha(1-\alpha)}\right|_{\rm max} &=& 
{2s_k(R^2-1)\over s_k^2+(R^2+1)^2}\,.
\end{eqnarray}

Using the results of section \ref{sec:results},
we find that the inclusion of the Lyman-$\alpha$ data 
significantly improves the 
previous bound on R to  $R<3$ at 95\% c.l. This bound comes
mainly from a combination of bounds on $2\beta\sqrt{\alpha(1-\alpha)}$
and $\niso$.

We did not find 
necessary to generate a $\ncor=0$ sampling for this model. In our 
results, the parameter $\delta_{\rm cor}$ has a flat distribution
and thus is unconstrained. We therefore expect similar results when
fixing it to zero.

\section{Conclusions}

In addition to CMB, LSS and SNIa data,
we used some recent Lyman-$\alpha$ forest data to further constrain
the bounds on possible CDM-isocurvature primordial fluctuations.
We find that the systematics induced $-$ in particular, those
associated with the recovery of
the linear dark matter power spectrum from the flux power
spectrum $-$ are greatly compensated by the valuable
information on the small-scale matter power spectrum
provided by the Lyman-$\alpha$ data. 

Before summarizing our results, it is worth mentioning
that when we omit the Lyman-$\alpha$ forest data our bounds
agree very well with those of Ref.~\cite{Kurki-Suonio:2004mn}.  The
authors of~\cite{Kurki-Suonio:2004mn} work with a pivot scale
$k_0=0.02~$Mpc$^{-1}$, but they also show how their results are modified
when they take $k_0=0.05~$Mpc$^{-1}$ like in the present paper: in that case
the agreement with us is particularly good. The comparison of our
results with the WMAP analysis from Ref.~\cite{Peiris} is more
puzzling: using or not some Lyman-$\alpha$ data, they always find much
stronger bounds on $f_{\rm iso}$ than us. It is true that we have one
more free parameter $\delta_{\rm cor}$, and that we do not introduce
a prior on the 2dF bias; however, even when we fix $\delta_{\rm
cor}=0$ and introduce such a prior, our $f_{\rm iso}$ bound remains
much weaker.  So far, private communications with the WMAP team did
not allow us to understand the origin of the discrepancy.

Using all our data set, we find at the 95\% confidence level,
an isocurvature fraction
$\alpha < 0.4$, a cross-correlation amplitude
$2 \beta \sqrt{\alpha(1-\alpha)}=0.1\pm0.4$, and an isocurvature
tilt $\niso=1.9 \pm 1.0$. The tilt of the correlation angle
remains unconstrained. If we switch to the basis used for instance in
Ref.~\cite{Peiris} we find $f_{\rm iso} < 0.75$ at 95\% c.l.

In the case of a curvaton scenario where CDM-creation occurs at the
decay of the curvaton -- a case in which the adiabatic and
isocurvature modes are maximally anti-correlated, $\beta=-1$, and
$\nad=\niso$ -- we find $f_{\rm iso} < 0.05$, still at the 95\%
confidence level. This requires that the fraction $r$ of the total
density in the curvaton field at that time be fine-tuned between 0.98
and one.  Finally, if we assume a double-inflation model with two
massive inflatons coupled only gravitationally, such that the heaviest
field decays into CDM, while the lightest one into standard model
particles, we find that the
mass ratio should obey $R < 3$ (95\% c.l.).

\section*{Acknowledgements.}
The
simulations were done at the UK National Cosmology Supercomputer
Center funded by PPARC, HEFCE and Silicon Graphics / Cray Research. MV
thanks PPARC for financial support. MB thanks the group at Sussex
University for their warm hospitality and acknowledges support by the
European Community programme HUMAN POTENTIAL under Contract
No. HPMT-CT-2000-00096. This work was supported in part by a
CICYT project FPA2003-04597, as well as a Spanish-French Collaborative
Grant between CICYT and IN2P3.
We whish to thank A. Liddle, H. Peiris and L. Verde for useful discussions.

\end{document}